\documentstyle[12pt,openbib]{article}
\title{Temperature and Relativity}
\author{Sandro S. Costa and George\ E.\ A.\ Matsas\\
        Instituto de F\'\i sica Te\'orica/UNESP \\
        Rua Pamplona 145\\
        01405-900-S\~ao Paulo, S\~ao Paulo\\
        Brazil}

\begin{document}
\maketitle

\begin{abstract}
We investigate whether inertial thermometers
moving in a thermal bath behave as being hotter or colder.
This question is directly related to the classical controversy
concerning how temperature transforms under Lorentz
transformations.  Rather than basing our arguments on
thermodynamical hypotheses, we perform straightforward
calculations in the context of relativistic quantum field theory.
For this purpose we use Unruh-DeWitt
detectors, since they have been shown to be reliable
thermometers in semi-classical gravity.  We
believe that our discussion helps in
definitely clarifying this issue.
\end{abstract}
\newpage

The problem of constructing a relativistic thermodynamical
theory was the source of an intense discussion for a long time (see
the editorials in Ref. \cite{EDITORIALS}, and Refs. [2-14]).
In particular, the question of how
temperature transforms under Lorentz transformations led some
distinguished physicists to reach exactly the opposite
conclusion of other equally distinguished ones. In order to set
up the problem clearly, consider a thermal bath with
temperature $T_0$ with
respect to its inertial rest frame $S_0$.
According to Einstein \cite{E}, Planck
\cite{P}, Tolman \cite{T}, Pauli \cite{Pau}, and Von Laue \cite{VL}
among others, a distinct inertial reference frame $S$
moving with constant velocity $v$ with respect to $S_0$ would ascribe to
this thermal bath a smaller temperature given by $T=T_0
\sqrt{1-v^2}.$ Lately, however, this result was challenged by
various authors.  Ott \cite{O} and Arzeli\`es
\cite{A}, for instance, reached exactly the opposite conclusion,
{\em i.e.} $T=T_0/ \sqrt{1-v^2}$, while Landsberg \cite{L}
claimed that $T=T_0$. Although differing in the results,
the approaches on which these discussions were based shared
the same thermodynamical nature.

This paper  is dedicated to  revisit this problem from a
completely different point of view. We shall investigate
explicitly  how a thermometer moving with constant velocity $v$
with respect to a thermal bath behaves by using quantum field
theory methods.  For this purpose, we use an Unruh-DeWitt
detector, since it has been shown to be a reliable thermometer
in the context of semi-classical gravity.  It is known that the
Minkowski vacuum is invariant under Lorentz transformations, and
consequently every inertial thermometer moving in the Minkowski
vacuum measures zero temperature. However, in the mid
seventies Unruh \cite{U} and Davies \cite{D} showed that a
uniformly accelerated thermometer in the Minkowski vacuum would
measure a temperature proportional to its proper acceleration.
Thus, it is natural to investigate in this context how an
inertial thermometer moving with respect to an ordinary thermal
bath would behave.  The main virtues of this approach is that it
is intrinsically covariant, and that it does not depend on any
thermodynamical hypotheses. Natural units ($c=k=\hbar=1$) will
be assumed throughout the paper.

An Unruh-DeWitt detector \cite{U,DW} is a two--level  monopole
which can be either in the ground state $\vert E_0 \rangle$ or
in an excited state $\vert E \rangle$.  For the sake of simplicity,
we couple the detector to massless scalar particles rather than
photons, since both fields behave similarly.  The total
excitation probability associated with a
detector moving  through a background
thermal bath with temperature $T_0$ can be
computed by standard quantum field theory methods \cite{BD}:
\begin{equation}
{\cal P}^{\rm exc} = c_0^2 \int_{-\infty}^{+\infty} d\tau
           \int_{-\infty}^{+\infty} d\tau'
           \; e^{-i\Delta E(\tau - \tau ')}
           G^+_{T_0} [x^\mu (\tau ), x^\mu (\tau ' )] ,
\label{P1}
\end{equation}
where $c_0$ is a small coupling constant between the detector
and the scalar field, $\Delta E \equiv  E-E_0$, and
\begin{equation}
G^+_{T_0} [x^\mu (\tau ), x^\mu (\tau ' )] =
-\sum_{n=-\infty }^{+\infty }
          \frac{(4\pi^2)^{-1}}{
          (t - t'-in/T_0  -i\varepsilon )^2 - \vert {\bf x}
	  - {\bf x}'\vert^2} ,
\label{Sum1}
\end{equation}
is the thermal Wightman function, where
$x^\mu (\tau )$ is the detector's world line, and
$\tau$ is its proper time. (We have assumed $\langle E \vert m(0)
\vert E_0 \rangle \equiv 1$, since the selectivity
only depends on the properties of the detector, and it is a
constant which always can be factored out \cite{BD}.)
Substituting the world line of a detector moving
with constant velocity $v$:
$t=\tau /\sqrt{1-v^2},$
$z= vt,$ $x=y=0$, in (\ref{Sum1}), we obtain
from (\ref{P1}) after some algebra \cite{GR} that
the detector's excitation rate is
\begin{equation}
\frac{d{\cal P}^{\rm exc}}{d\tau} = c_0^2 \frac{
T_0 \sqrt{1-v^2}}{4 \pi v} \ln\left[
         \frac{1-e^{- \Delta E \sqrt{1+v}/T_0 \sqrt{1-v}}}
         {1-e^{- \Delta E \sqrt{1-v}/T_0 \sqrt{1+v}}}
   \right].
\label{Crucial}
\end{equation}
Here it is enough to consider the detector as being
permanently switched on, rather than as being switched off
asymptotically \cite{HMP}, since this is a
stationary situation. In the limit $v \to 0$ we obtain the usual
black body excitation rate \cite{BD}
\begin{equation}
\frac{d{\cal P}^{\rm exc}}{d\tau} = \frac{c_0^2
\Delta E}{2 \pi (e^{\Delta E/T_0} -1 )} ,
\end{equation}
while in the limit $v \to 1$,
we obtain $d{\cal P}^{\rm exc}/d\tau \to 0$.
This means that ultra-relativistic detectors
do not interact appreciably with
the background thermal bath (see Fig.1).
It can be understood physically on energy conservation
grounds. An inertial Unruh-DeWitt detector only responds  to the
presence of modes with very precise frequency $\omega = \Delta
E$ as measured in its rest frame $S$. Hence, when $v$ is large,
 half of the particles present in
the  bath are so much red-shifted while the other half are so much
blue-shifted that it cannot be excited.

It is clear from (\ref{Crucial}) that the moving detector does
not respond according to a black body spectrum. The particle
number distribution $n(\omega)$ in the frame $S$
can be written directly from (\ref{Crucial})  as
\begin{equation}
n(\omega ) d^3{\bf \tilde k}  = \frac{
T_0 \sqrt{1-v^2} }{4 \pi v } \ln\left[
         \frac{1-e^{- \omega \sqrt{1+v}/T_0 \sqrt{1-v}}}
         {1-e^{- \omega \sqrt{1-v}/T_0\sqrt{1+v}}}
   \right] d^3{\bf \tilde k} ,
\label{Crucial2}
\end{equation}
where $d^3{\bf \tilde k} \equiv d^3{\bf k}/4 \pi^2 \omega =
d\omega \omega/\pi$.
Notice that in the limit $v \to 0$, we obtain the usual black
body particle number distribution
\begin{equation}
n_0 (\omega ) d^3{\bf \tilde k} =
\frac{\omega^2 d\omega }{2 \pi^2 (e^{\omega /T_0}-1)} .
\label{Pl}
\end{equation}

Let us analyze the infrared sector, $\omega << T_0$,
of $n(\omega )$ for $v<<1$.
This sector can be physically
probed through a slow moving two--level
detector with $\Delta E << T_0$.
In this case, we obtain from (\ref{Crucial2})
\begin{equation}
n(\omega << T_0) \approx T_0 (1
-v^2/6) /2 \pi .
\label{A}
\end{equation}
Analogously, we obtain from (\ref{Pl}) in the region $\omega <<T_0$
\begin{equation}
n_0(\omega <<T_0) \approx T_0 /2\pi .
\label{B}
\end{equation}
Now, comparing (\ref{A}) and (\ref{B})
we are able to define naturally an effective temperature
of the thermal bath as measured in $S$ by
\begin{equation}
T \approx T_0 (1 -v^2/6) .
\label{FANT}
\end{equation}
It is now instructive to compare this result against  the one obtained
by completely different means in Refs. [11-13],
and recently corroborated thermodynamically
in \cite{AG}.  Using special relativity,
Bracewell and Conklin, Peebles and Wilkinson,
and Henry et al, showed that an observer in S looking at some
{\em fixed} direction would still map out the background thermal bath into
a blackbody spectrum.
Thus, they defined in the moving frame an effective {\em directional}
temperature
\begin{equation}
T(T_0,v,\theta) = \frac{T_0 \sqrt{1-v^2}}{1- v \cos \theta},
\label{TVTETA}
\end{equation}
as a function of the angle $\theta$ as measured in $S$
between the axis of motion and the
direction of observation. Since our detector
is a monopole, the best we can do is to compare our results
with the average of (\ref{TVTETA}) in the solid angle
\begin{equation}
\langle T \rangle  = \frac{1}{4\pi } \int  T(T_0, v, \theta)
d\Omega .
\end{equation}
Performing this integral we obtain
\begin{equation}
\langle T \rangle = T_0 (1-v^2/6),
\end{equation}
which coincides with (\ref{FANT}) .

The results above suggest that a thermometer moving with respect
to a background thermal bath would {\em always} ascribe  a
smaller temperature in comparison with another thermometer lying
at rest in the bath. Nevertheless, we are not allowed to
make such a general claim. In order to define uniquely an
effective temperature  in the moving frame $S$,
we should be able first to
express (\ref{Crucial2}) in the black body  form [see Eq. (\ref{Pl})]
\begin{equation}
n( \omega) d^3{\bf \tilde k} =
\frac{\omega^2 d\omega}{2\pi^2 (e^{ \omega /T }-1)}  ,
\label{Plm}
\end{equation}
for some analytic function $T=T(T_0,v)$ without any angular dependence.
Since this is
impossible, it is necessary to define
some prescription to
generalize the concept of temperature as above. Notwithstanding, we
emphasize that
different prescriptions may result in opposite conclusions.

In order to illustrate how a thermometer sensitive to the whole
spectrum could reach a distinct conclusion in
comparison with (\ref{FANT}), let us consider a device sensitive
to the whole particle spectrum rather than just to the
low-frequency part.
The particle density associated with some particle distribution
$n (\omega )$ can be calculated by
\begin{equation}
\bar n =  \int  n (\omega) d^3{\bf \tilde k} .
\label{SB}
\end{equation}
In order to obtain the particle density in $S_0$, we use
(\ref{Pl}) in (\ref{SB}) obtaining \cite{Pa}
\begin{equation}
\bar n_0 = \frac{T_0^3 \zeta (3)}{\pi^2} ,
\label{po0}
\end{equation}
where $\zeta (x)$ is the zeta function.
Analogously, in order to calculate the particle density in $S$, we
use  (\ref{Crucial2}) in (\ref{SB}), obtaining
\begin{equation}
\bar n = \frac{\gamma T_0^3 \zeta (3)}{\pi^2} ,
\label{po}
\end{equation}
where $\gamma \equiv (1-v^2)^{-1/2}$.
Notice that $\bar n_0$ and $\bar n$ only differ
by a $\gamma$ factor which
expresses the Lorentz contraction of the volume.
By comparing (\ref{po0}) and (\ref{po}), it
is natural according to this procedure
to define another effective temperature for the bath as
measured in $S$ given by
\begin{equation}
T= T_0 (1-v^2)^{-1/6} .
\label{FANT2}
\end{equation}
In the limit $v<<1$, (\ref{FANT2}) can be cast in the simpler form
\begin{equation}
T \approx T_0 (1 + v^2/6).
\label{FANT3}
\end{equation}

Hence, a slow moving thermometer sensitive only to the low-energy
part of the spectrum would  measure
according to (\ref{FANT}) an
effective temperature $T<T_0$, while a slow moving device sensitive
to the whole energy spectrum could measure
according to (\ref{FANT3}) an effective temperature $T>T_0$.
This is so because in the last case the high-frequency part of
the spectrum plays a significant role in increasing the
temperature (see Fig.1).
There is no contradiction between these results,
since they do not disagree concerning any real events. On
the contrary, both are a direct consequence of (\ref{Crucial}).

In summary, we have  investigated the classical
controversy about whether moving thermometers in a background
thermal bath with temperature $T_0$ behave as being
hotter or colder. For this
purpose we have used an Unruh-DeWitt detector as a
thermometer, since it had been introduced with
success in the context of
semi-classical gravity in connection with black hole evaporation.
We have obtained that slow
moving thermometers sensitive only to the infrared part of
the spectrum would measure a temperature $T \approx T_0 (1-v^2/6)$.
This result is compatible with the one obtained in [11--13]
by completely different means. However, other devices sensitive to
the whole spectrum may give different results (\ref{FANT3}).
We have argued that this result rather than expressing a
contradiction, just reflects the fact that
the frequency spectrum in
the moving frame is not the usual black body one.
It is important to recall that
inequivalent definitions of temperature can be allowed provided
they do not disagree over any real events. However, this warns us
that the question about how temperature transforms under Lorentz
transformations does not make sense unless one defines carefully what
the considered thermometer is.

\begin{flushleft}
{\bf{\large Acknowledgements}}
\end{flushleft}

We are indebted to stimulating discussions
with R. Aldrovandi, and S. Salinas. We are also grateful to
R. Aldrovandi, J.S. Guill\'en, D. Hadjimichef, U. Heller, A.
Natale, and V. Pleitez for their comments in the final
manuscript.  (SC) was full supported by Coordenadoria de
Aperfei\c coamento de Pessoal de N\'\i vel Superior, while (GM)
was partially supported by  Conselho Nacional de Desenvolvimento
Cient\'\i fico e Tecnologico.

\newpage

\begin{figure}
\protect
\caption{{$ d{\cal P}^{\rm exc}/d\tau $ is plotted
as a function of the detector's velocity $v$, and its energy gap
$\Delta E$ for $T_0 =1$, and $c_0=10^{-1}$.
For $v=0$ we obtain the usual black body spectrum, while for
$v \to 1$ the excitation rate vanishes. For detectors with $\Delta
E<< T_0$ the maximum excitation rate is obtained for $v=0$, while
for detectors with $\Delta E >> T_0$ the maximum excitation rate
is obtained for some value $v > 0$. This is the reason why
slow moving thermometers which only probe the infrared part of
the spectrum measure a temperature $T<T_0$, while thermometers which
probe the whole spectrum may measure a temperature $T>T_0$. }}
\label{fig:iner}
\end{figure}

\end{document}